\documentclass[apjl]{emulateapj}
\providecommand\scription[2]{\scriptsize#1$\;${\scriptsize\uppercase\expandafter{\romannumeral #2}}\relax}%
\providecommand\tabion[2]{#1$\;${\tiny\uppercase\expandafter{\romannumeral #2}}\relax}%
\providecommand{\CC}{Complex~C}
\providecommand{\degree}{\ensuremath{^{\circ}}}

\providecommand{\Msun}{\,\ensuremath{\mbox{M}_{\odot}}}
\providecommand{\Msunyr}{\,\ensuremath{\mbox{M}_{\odot}\,\mbox{yr}^{-1}}}
\providecommand{\kms}{\,\ensuremath{\rm{km\,s}^{-1}}}
\providecommand{\cm}{\,\ensuremath{\mbox{cm}^{-2}}}
\providecommand{\A}{\,\ensuremath{\mbox{\AA}}}
\providecommand{\mA}{\,\ensuremath{\mbox{m\AA}}}

\providecommand{\Jykms}{\,\ensuremath{\mbox{Jy\,km\,s$^{-1}$}}}

\providecommand{\kpc}{\,\ensuremath{\mbox{kpc}}}

\providecommand{\arcmin}{\mbox{$^\prime$}}%
\providecommand{\arcsec}{\mbox{$^{\prime\prime}$}}%

\providecommand{\Zsun}{\,\ensuremath{Z_{\odot}}}
\providecommand{\Myr}{\,\ensuremath{\mbox{Myr}}}
\providecommand{\Gyr}{\,\ensuremath{\mbox{Gyr}}}
\providecommand{\HI}{\ensuremath{\mbox{\ion{H}{1}}}}

\providecommand{\Halpha}{\ensuremath{\mbox{H}\alpha}}

\providecommand{\He}{\ensuremath{\mbox{H}\epsilon}}
\providecommand{\Lya}{\ensuremath{\mbox{Ly\,}\alpha}}

\providecommand{\CaII}{\ensuremath{\mbox{\ion{Ca}{2}}}}
\providecommand{\CaK}{\ensuremath{\mbox{\ion{Ca}{2}~K}}}
\providecommand{\CaH}{\ensuremath{\mbox{\ion{Ca}{2}~H}}}
\providecommand{\CaHK}{\ensuremath{\mbox{\ion{Ca}{2}~H \& K}}}

\providecommand{\NaD}{\ensuremath{\mbox{\ion{Na}{1}~D}}}

\providecommand{\NaDb}{\ensuremath{\mbox{\ion{Na}{1}~D$_{2}$}}}

\providecommand{\MHI}{\ensuremath{\mbox{M$_{\scHI}$}}}
\providecommand{\ACa}{\ensuremath{\mbox{A(\CaII)}}}
\providecommand{\scHI}{\ensuremath{\mbox{\scription{H}{1}}}}

\providecommand{\vLSR}{\ensuremath{v_{\mbox{\tiny LSR}}}}

\providecommand{\vdev}{\ensuremath{v_{\mbox{\tiny dev}}}}
\providecommand{\SiIV}{\ensuremath{\mbox{\ion{Si}{4}}}}
\providecommand{\OVI}{\ensuremath{\mbox{\ion{O}{6}}}}
\providecommand{\CIV}{\ensuremath{\mbox{\ion{C}{4}}}}
\providecommand{\NV}{\ensuremath{\mbox{\ion{N}{5}}}}
\providecommand{\FeH}{\ensuremath{\mbox{[Fe/H]}}}
\providecommand{\Teff}{\ensuremath{\mbox{T}_{\mbox{\scriptsize eff}}}}
\providecommand{\logg}{\ensuremath{\mbox{log\,{\it g}}}}
\providecommand{\etal}{\ensuremath{\mbox{et~al.}}}
\providecommand{\vhelio}{\ensuremath{v_{\mbox{\tiny helio}}}}
\providecommand{\OI}{\ensuremath{\mbox{\ion{O}{1}}}}
\providecommand{\MgII}{\ensuremath{\mbox{\ion{Mg}{2}}}}
\providecommand{\NHI}{\ensuremath{N(\HI)}}
\providecommand{\NCa}{\ensuremath{N(\CaII)}}
\providecommand{\NCaNHI}{\ensuremath{N(\CaII)/N(\HI)}}

\providecommand{\logNHI}{\ensuremath{\mbox{log\,N(\HI)}}}
\providecommand{\TiII}{\ensuremath{\mbox{\ion{Ti}{2}}}}
\def\spose#1{\hbox to 0pt{#1\hss}}
\def\simlt{\mathrel{\spose{\lower 3pt\hbox{$\mathchar"218$}}
     \raise 2.0pt\hbox{$\mathchar"13C$}}}
\def\simgt{\mathrel{\spose{\lower 3pt\hbox{$\mathchar"218$}}
     \raise 2.0pt\hbox{$\mathchar"13E$}}}
\providecommand{\EW}{\ensuremath{\mbox{W}_{\lambda}}}
\providecommand{\sigEW}{\ensuremath{\sigma(\mbox{W}_{\lambda})}}
\providecommand{\CCMtot}{\ensuremath{M_{\rm tot} = 8.2^{+4.6}_{-2.6} \times 10^{6}\Msun}}
\providecommand{\CCMHI}{\ensuremath{\MHI\ = 4.9^{+2.8}_{-2.2} \times 10^6 \Msun}}

\shorttitle{The Distance to Complex C}
\shortauthors{Thom et al.}

\begin{document}

\title{An Accurate Distance to High-Velocity Cloud Complex C}


\author{
C.~Thom\altaffilmark{1,2}, 
J.E.G.~Peek\altaffilmark{3}, 
M.E.~Putman\altaffilmark{4}, 
Carl~Heiles\altaffilmark{3}, 
K.M.G.~Peek\altaffilmark{3}, 
R.~Wilhelm\altaffilmark{5}
}

\email{cthom@uchicago.edu, goldston@astro.berkeley.edu, mputman@umich.edu, heiles@astro.berkeley.edu, kpeek@astron.Berkeley.edu, ron.wilhelm@ttu.edu}

\altaffiltext{1}{Department of Astronomy and Astrophysics, University of Chicago, Chicago, IL, 60637, USA}
\altaffiltext{2}{Kali Institute for Cosmological Physics, University of Chicago, Chicago, IL, 60637, USA}
\altaffiltext{3}{Astronomy Department, University of California, Berkeley CA, 94720, USA}
\altaffiltext{4}{Department of Astronomy, University of Michigan, Ann Arbor, MI 48109, USA}
\altaffiltext{5}{Department of Physics, Texas Tech University, Lubbock, TX 79409, USA}


\begin{abstract}

  We report an accurate distance of $d = 10\pm2.5\kpc$ to the high-velocity cloud \CC. Using high
  signal-to-noise Keck/HIRES spectra of two horizontal-branch stars, we have detected \CaK\
  absorption lines from the cloud.  Significant non-detections toward a further 3 stars yield robust
  lower distance limits. The resulting \HI\ mass of \CC\ is \CCMHI; a total mass of \CCMtot\ is
  implied, after corrections for helium and ionization. At 10\kpc, \CC\ has physical dimensions $3
  \times 15\kpc$, and if it is as thick as it is wide, then the average density is $log\,\langle n
  \rangle \simeq -2.5$. We estimate the contribution of \CC\ to the mass influx may be as high as
  $\sim0.14\Msunyr$.

\end{abstract}


\section{Introduction}

The halo of the Milky Way contains clouds of neutral hydrogen (\HI) gas representing the flow of
baryons into, and out of, the Galactic disk. Identified by their velocities, these high-velocity
clouds (HVCs)\footnote{HVCs are defined as having $|\vLSR| > 90\kms$ and hence do not co-rotate with
  the Galaxy.}  have been observed in \HI\ 21cm emission for more than 40 years
\citep{muller-etal-63-HVC-discovery}. Since their distances cannot be determined by the application
of a kinematic model, the mass scale of the flow is uncertain and the clouds' origin and impact on
the disk open to speculation.

Since their discovery, many explanations have been offered for the HVC phenomenon, with a
corresponding range of distances; most of these explanation can be traced to
\citet{oort-66-HVC-origins}.  One possible origin for some HVCs is a supernovae-driven Galactic
fountain, in which gas is expelled from the disk into the halo, condenses and then returns in a
cyclic flow \citep{bregman-80, houck-bregman-90}. \citet{gibson-etal-01-ComplexC} suggested this
model may be appropriate for \CC\ based on observed abundance variations across the complex. In
other galaxies, there is evidence that some (but not all) of the extra-planar \HI\ observed in
edge-on spiral NGC\,891 originates in such a fountain \citep{oosterloo-etal-07-NGC891}. Theoretical
models suggest fountain gas may rise as high as $\sim10\kpc$ above the disk
\citep{deAvillez-00-galfountain}. Dwarf galaxy accretion is another avenue for mass input; the
well-known Magellanic Stream \citep{mathewson-etal-74-MS, putman-etal-03-mshvc} is the canonical
example, but this mechanism has been suggested for other MW satellites
\citep{putman-etal-04-sgr}. Another recent proposal posits the clouds to be condensing out of the
hot MW halo \citep{maller-bullock-04, kaufmann-etal-06-condensation-simulation, sommer-larsen-06,
  peek-etal-07-accretion}, at a range of distances up to $\sim150\kpc$.
\citet{peek-etal-07-accretion} suggested that HVC masses are inversely proportional to their
distance, reasoning that condensing clouds are small at large distance, and form the seeds of the
large HVC complexes, into which they grow via accretion as they approach the disk. If this scenario
is correct, then all the large HVC complexes would be close ($d \sim10-15\kpc$).

\CC\ was first mapped by \citet{hulsbosch-raimond-66-survey}. Its large angular size ($\sim1600
sq.~deg.$) means that a number of UV-bright quasars align with high column-density gas, allowing
metallicity determinations by absorption line spectroscopy.  Based on a metallicity measurement of
$\sim0.1\,Z_{\odot}$, \citet{wakker-etal-99-nature} suggested \CC\ is a low-metallicity, infalling
cloud, fueling Galaxy formation. \citet{gibson-etal-01-ComplexC} later reported metallicities as
high at $0.3\,Z_{\odot}$, proposing that \CC\ may instead a be product of star formation in the
Galaxy or a disrupted satellite. Most recently \citet{collins-etal-07-ComplexC} confirmed this
evidence of abundance variations across the cloud independent of ionization effects, suggesting the
possible mixing of primordial and enriched gas \citep[see also][]{tripp-etal-03-ComplexC}.

\begin{deluxetable*}{lrrrrrrr}[b]
\tabletypesize{\scriptsize}
\tablecaption{\label{tab: basic-info}Journal of Observations}
\tablewidth{0pt}
\tablehead{
\colhead{Target} &
\colhead{$l$} &
\colhead{$b$} &
\colhead{$g_0$} &
\colhead{S/N} &
\colhead{\vhelio} &
\colhead{Dist} &
\colhead{Label} \\
\colhead{} &
\colhead{(deg)} &
\colhead{(deg)} &
\colhead{(mag)} &
\colhead{} &
\colhead{(\kms)} &
\colhead{(\kpc)} &
\colhead{} 
}

\startdata
SDSS J173424.01+601735.3 &   89.1 & 32.8 &  15.9 &  60 & $-482.6 \pm 1.4$ & $12.8 \pm 3.2$ & S135 \\
SDSS J172009.78+612502.3 &   90.6 & 34.4 &  14.7 & 100 & $-322.3 \pm 3.3$ & $ 7.8 \pm 2.0$ & S139 \\
SDSS J150335.53+623513.5 &  100.7 & 48.4 &  15.6 &  60 & $-201.2 \pm 2.5$ & $11.3 \pm 2.8$ & S437 \\
SDSS J153915.24+575731.7 &   91.2 & 47.5 &  15.9 &  80 & $ -14.4 \pm 1.5$ & $10.2 \pm 2.6$ & S441 \\
SDSS J133654.82+622241.5 &  114.0 & 54.0 &  15.4 &  65 & $  26.5 \pm 0.9$ & $10.4 \pm 2.6$ & S674 \\
\enddata

\tablecomments{SDSS names are created from the RA and DEC (J2000) of the object, truncating
  co-ordinates. Galactic co-ordinates have been rounded to 0.1\degree. Magnitudes are extinction
  corrected $g$-band, taken from the SDSS database. Signal-to-noise measures are given for the
  continuum region near \CaK. For convenience, we label each star with a number, corresponding to
  its numerical position in the catalogue of \citet{sirko-etal-04a}.}
\end{deluxetable*}

There is no convincing evidence of dust in \CC\ \citep[e.g.][]{collins-etal-03}, nor any
detections of molecular hydrogen \citep{murphy-etal-00-ComplexC-FUSE, richter-etal-01-MolecularGas}.
Observations with the Wisconsin \Halpha\ Mapper (WHAM) have detected \Halpha\ emission at the same
velocities as the \HI\ emission \citep{tufte-etal-98-WHAM, haffner-etal-03-WHAM-survey}. Higher
ionization species, such as \CIV, \SiIV, \NV\ and \OVI\ have been seen in absorption toward QSOs
\citep{sembach-etal-03-OVI-HVC, fox-etal-04-ComplexC}.  \citet{fox-etal-04-ComplexC} studied the
\OVI\ and other highly ionized species, concluding that they are most likely to arise in conductive
or turbulent interface regions between the warm neutral gas seen in \HI, and the surrounding hot
halo medium.

HVCs are often invoked by Galactic chemical evolution models as a source of infalling star formation
fuel, which is needed in order to reproduce the metallicity distribution function of stars in the
solar neighborhood \citep[the ``G Dwarf'' problem; e.g.][]{pagel-patchett-75,
  alibes-etal-01-infall-rate}.  As the largest HVC on the sky, \CC\ could potentially contribute a
significant fraction of the Galaxy's future star formation fuel. For HI clouds, the mass is related
to the distance by $\MHI = 0.236\,S\,d^2$, where $S$ is the total \HI\ flux observed (in \Jykms) and
$d$ the distance in \kpc. For \CC, $S = 2\times10^5\Jykms$ \citep{wakker-vanwoerden-91-catalogue}.
Recently, \citet{wakker-etal-07-ComplexC} reported a firm distance limit to \CC\ of $3.7-11.2\kpc$,
with a weaker limit of $6.7-11.2\kpc$. Here we employ high-quality Keck spectra to determine the
most accurate measurement of $d$ for \CC\ to date.

\section{Data}
\label{sec: data}

Five horizontal branch stars from the catalogue of \citet{sirko-etal-04a} that align with Complex~C
were observed with the High Resolution Echelle Spectrometer \citep[HIRES][]{vogt-etal-94-HIRES} on
the Keck I telescope on 08 June 2007 and 09 June 2007 (UT).  We employed the UV cross-disperser to
maximize throughput at \CaK\ ($\lambda\,3933.663\A$), and binned the data $2\times$ in the spatial
direction. In this configuration, the \NaD\ lines are within our wavelength coverage, but are at the
extreme edge of the order.  The $7\arcsec \times 0.8\arcsec$ slit yielded a resolution of $R
\sim48,000$. Over the course of the two nights, the seeing varied from $\sim1.2 - 1.8\arcsec$, but
we nevertheless obtained excellent quality data. Table~\ref{tab: basic-info} summarizes our targets.
For these stars, stellar parameters were obtained using the techniques described in
\citet{wilhelm-etal-99a-classification}. Absolute magnitudes and distances for the stars were
determined by comparing \Teff, \logg, and \FeH\ to the run of theoretical isochrones of
\citet{girardi-etal-04-isochrones} and \citet{dorman-92-HB-isochrones}; distances are given in the
text, and in Table~\ref{tab: summary}. External tests of the stellar technique against stars in
globular clusters indicate that distances are accurate to $\sim25\%$ (Wilhelm \etal\ 2008, in
prep.). We thus adopt a 25\% error as our formal distance error; in all cases this error is greater
than that estimated from the stellar classification, sometimes significantly. Continuing calibration
efforts of the Segue Stellar Parameter Pipeline \citep[][]{lee-etal-08-sspp-I} should improve this
accuracy in the future.

To reduce the data, we used the HIRES Redux pipeline (Bernstein, Burles \& Prochaska, 2008, in
prep.) which is distributed as part of the {\it xidl} package\footnote{http://www.ucolick.org/\~{
  }xavier/IDL/index.html}. The package applies standard bias and flat-field corrections, performs a
2-D wavelength calibration, and extracts the individual orders to give final 1-D spectra and
associated errors.  For equivalent width measurements, we fit the local continuum with a low-order
Legendre polynomial and directly integrate the data, determining error contributions from both the
noise and continuum fitting \citep{sembach-savage-92-EW}. In general, the HVC \CaH\
($\lambda\,3968.467\A$) line is in the wings (and in some cases in the core) of the broad stellar \He\
($\lambda\,3970.072\A$) line. The \He\ line is sufficiently broad that it completely dominates the
echelle order, and the true continuum cannot accurately be determined. Since we care only about the
HVC \CaH\ line, we fit a pseudo-continuum along the balmer line wing where possible, removing its
contribution to the absorption.

\HI\ spectra were primarily drawn from the combined Leiden/Argentine/Bonn (LAB) survey, which has a
beam-width of $\sim$36\arcmin\ (FWHM) \citep{kalberla-etal-05-LAB}. In one case (S441---see
Table~\ref{tab: basic-info}), an \HI\ spectrum from Effelsberg (9\arcmin\ beam) is
available. Heliocentric radial velocities (RVs) for all 5 stars were determined by fitting the
positions of unblended metal lines in the echelle data. These are also given in Table~\ref{tab:
  basic-info}. In most cases $\sim30$ or more lines were available, but for S139, only 12 unblended
metal lines are available in our Keck spectrum. Errors on RVs are the $1\sigma$ gaussian dispersion
of the ensemble of measured lines.

Using the absorption line technique, upper or lower distance limits are placed on the distance to
the HVC gas based on the detection or non-detections of absorption lines due to the gas, in the
spectrum of stars at known distance. The technique is described in detail elsewhere
\citep{schwarz-etal-95, thom-06-PhD-thesis}; Figure~1 of \citet{schwarz-etal-95} is particularly
enlightening. Since the stars must lie at known (or knowable) distances, horizontal-branch and
RR-Lyra stars are the most commonly used, but in principle any hot star at known distance may be
used; hot stars are desirable to reduce the number of metal lines in the stellar spectrum, which may
confuse the HVC absorption. In order to maximize the probability of optical detection, strong
resonance transitions are the most appropriate. In the optical, \CaHK\ are the most common, but the
\NaD\ lines may also be used \citep[see e.g.][]{thom-etal-06-complexWB}. Strong UV transitions
such as \OI\ have also been used \citep{danly-etal-93-ComplexM}, and with the scheduled installation
of the Cosmic Origins Spectrograph on the upcoming HST servicing mission, \MgII\ may also prove
useful.

\section{Results}

In the following subsections, we discuss the individual lines of sight first for the detections,
then for the non-detections. We consider the upper distance limits that are set by the detections,
and the lower distance limits that can be set by the non-detections. Finally we summarize our
results, both in tabular and pictorial form.  

\begin{figure}[t]
  \epsscale{1.2}
  \plotone{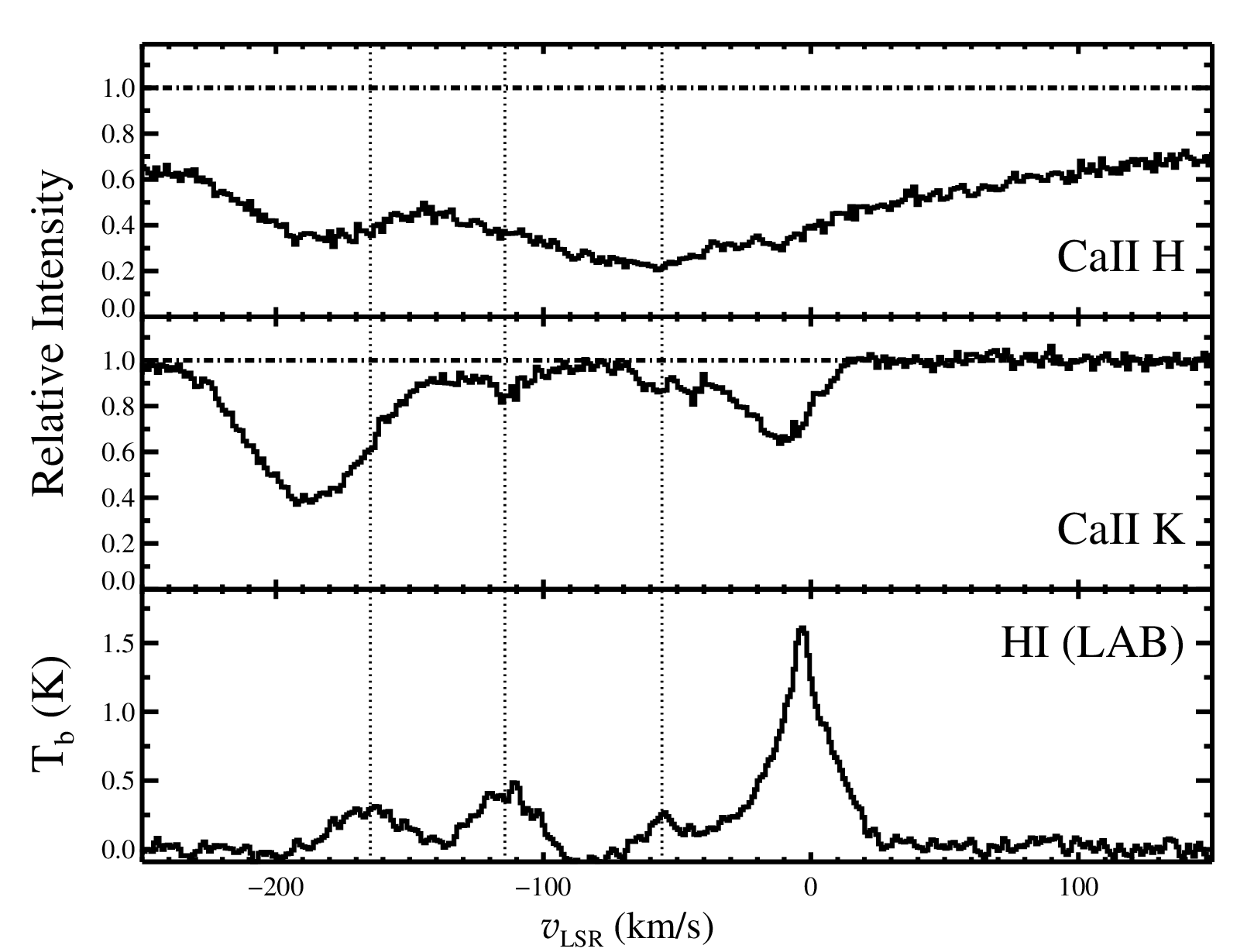}
  \caption{S437: Detection of Complex C in absorption toward S437. In this direction, we see the
    HVC in emission at $\vLSR = -115\kms$, which is clearly seen in \CaK\ absorption. There is also
    a higher velocity component seen in emission (sometimes called the ``high-velocity ridge'') at
    \vLSR\ = -165\kms. This higher-velocity component is not detected, since HVC \CaK\ absorption is
    obscured by the the stellar \CaK\ line, and the \CaH\ line is obscured by both core of the
    stellar \He\ line and the stellar \CaH\ absorption. IVC gas is evident at $\vLSR -56\kms$.}
  \label{fig: S437}
\end{figure}

\subsection{Detections}

S437, $d \sim11 \kpc$---Toward the star S437 two HVC emission components can be seen in the HI
spectrum available from the LAB survey (Figure~\ref{fig: S437}), at $\vLSR = -115\kms$ and $\vLSR =
-165\kms$, with corresponding column densities $\NHI = 2.1 \pm 0.3$ and $\NHI = 1.5 \pm 0.4$
($\times 10^{19}\cm$). The HVC emission component at $-115\kms$ is clearly evident in absorption at
\CaK, but the corresponding \CaH\ line is lost in the core of the \He\ balmer line. Since the HVC
feature falls in the very core of the \He\ line, we do not attempt to fit the continuum to the
balmer line wings. The HVC component at $-165\kms$ is lost in the strong stellar \CaII\ lines.
Emission and absorption at $\vLSR -56\kms$ from the intermediate-velocity cloud (IVC) known as the
``IV-arch'' can also be seen. This cloud is known to be nearby \citep[$0.4 < z < 3.5\kpc$;
][]{ryans-etal-97-IVArch-ComplexM}.

\begin{figure}[t]
  \epsscale{1.2}
  \plotone{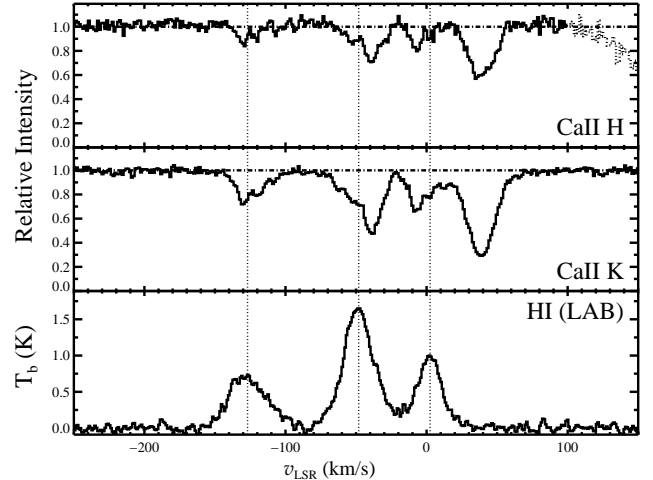}
  \caption{S674: Strong absorption from Complex~C is evident at -127\kms, setting an upper distance
    limit. Intermediate-velocity absorption from the IV-arch can also be seen at -48\kms. See text
    for further details. }
  \label{fig: S674}
\end{figure}


S674, $d \sim10\kpc$---The optical spectra of S674, shown in Figure~\ref{fig: S674}, provides a
second detection toward the high latitude part of \CC. Absorption is detected from \CC\ at $\vLSR =
-127\kms$ setting an upper distance limit. We also detect the nearby IV-arch at $\vLSR = -48\kms$ ,
and the MW disk in absorption. Note that we have fitted the \CaH\ continuum along the wing of the
stellar \He\ balmer line, removing it from Figure~\ref{fig: S674}. The fall-off in the continuum
red-ward of $\vLSR\ = 100\kms$ in the \CaH\ line is where the balmer line transitions from wing to
core; since we use only a low order polynomial, and do not attempt to match the entire balmer line
profile, the fit breaks down in this region.

We clearly detect multiple absorption components in the optical spectra which do not always align
well with the \HI\ emission positions. This is present in both the HVC absorption, as well as the
IVC and disk gas. Since the \HI\ data sample 36\arcmin\ on the sky, while the optical data trace a
pencil beam, this may be a consequence of beam smearing. An interferometer map would be required to
confirm this. Note, however, that in all cases, the optical absorption falls within the line limits
defined by the \HI\ data.

\subsection{Non-Detections}

\begin{figure}[t]
  \epsscale{1.2}
  \plotone{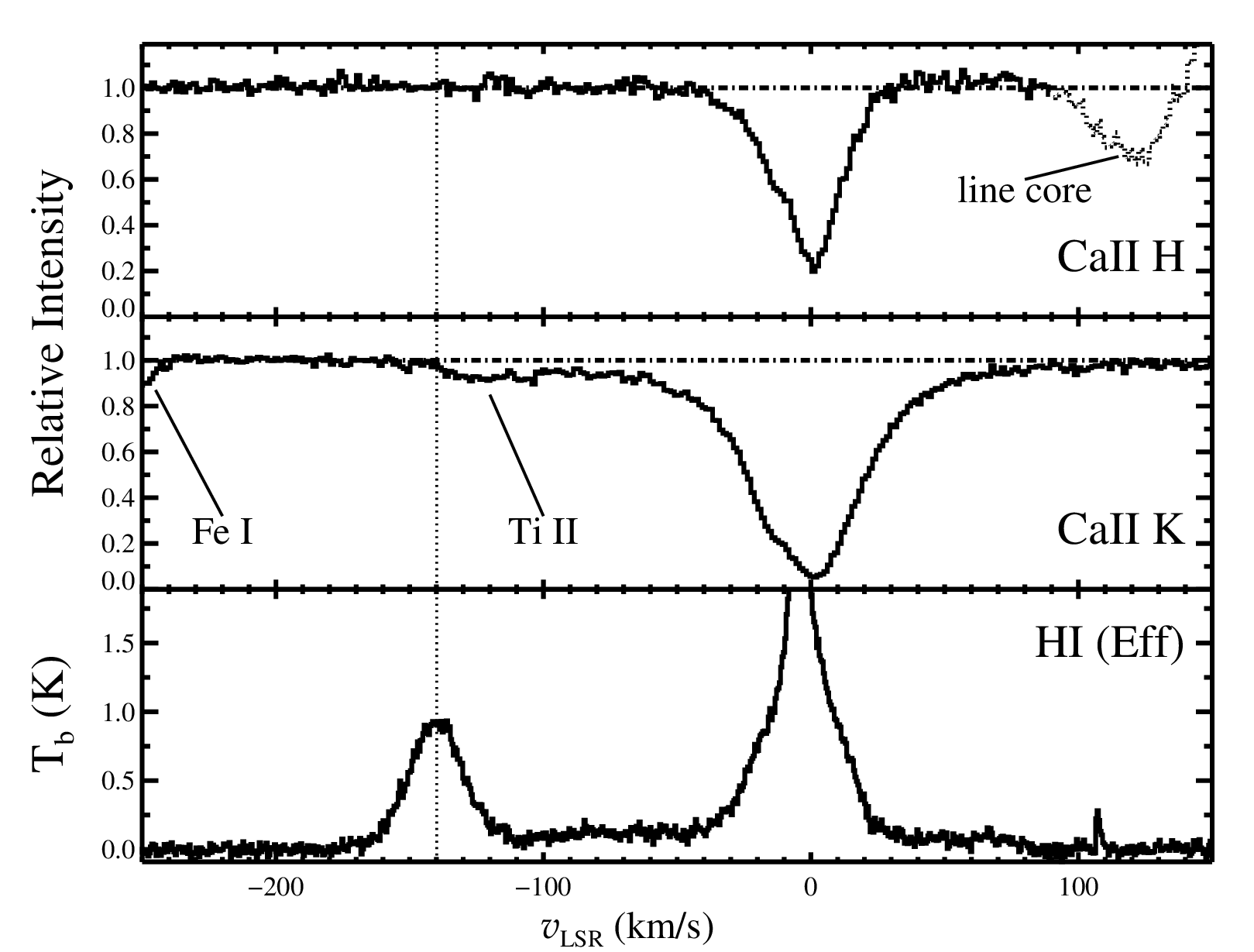}
  \caption{S441: No HVC absorption is present in the sightline toward S441. The weak feature in
    \CaK\ red-ward of the HVC position is stellar \TiII.  For the \CaH\ profile, we fit a
    pseudo-continuum along the broad wing of the \He\ balmer line. The dotted section in this top
    panel marks the core of the Balmer line, where the continuum fit is unreliable. }
  \label{fig: S441}
\end{figure}

S441, $d \sim10\kpc$---At roughly the same latitude as our two detections, and only $\sim28\arcmin$
from the Mrk\,290 sightline, this line of sight has the most complicated optical
spectrum. Figure~\ref{fig: S441} shows both the optical and radio spectra.  Note that for this
sightline, the \HI\ spectrum comes from the Effelsberg telescope, which has a substantially smaller
beam than the coarse resolution of the LAB survey.

The optical data in the \CaK\ region show an absorption line slightly to the red ($\vLSR =
-125\kms$) of where we expect the HVC absorption to lie ($\vLSR = -140\kms$) . This line is {\it
  not} HVC absorption, but rather a \TiII\ line at a rest wavelength of 3932.020\AA\
\citep{meggers-etal-75-NIST-Ti}.  This identification is made based on several factors. Firstly, in
the stellar rest frame, a fit to this line gives a rest wavelength 3932.00\AA. This differs from the
expected position by only 1.6\kms, in good agreement with the measured radial-velocity precision
(1.4\kms). Second, the strength of this line is consistent with other \TiII\ lines in the spectrum
of S441 with similar ionization potential and log\,{\it gf}. Finally, the strength of this line is
$\EW = 25.9 \pm 1.4\mA$. If this was an HVC \CaK\ absorption line that was offset from the \HI\
position, the corresponding \CaH\ line would be 13\mA. The noise level in the \CaH\ region [$\sigEW
= 2.5\mA$] is low enough that we would detect any HVC absorption at $5\sigma$ confidence. We can
thus exclude this possibility.

\begin{figure}[t]
  \epsscale{1.2}
  \plotone{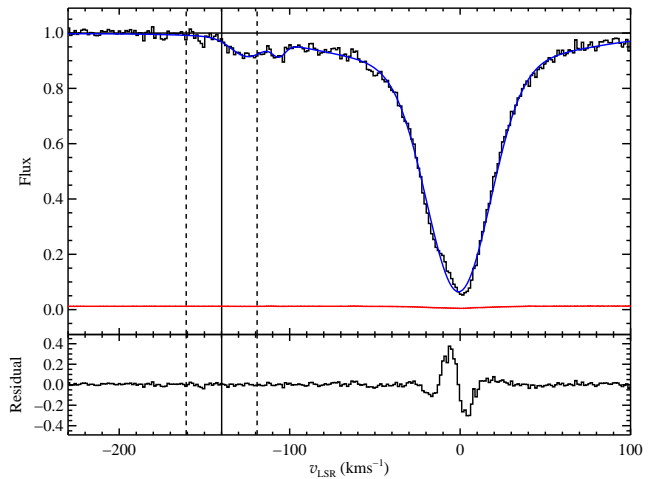}
  \caption{Fit to the stellar absorption in S441. The solid vertical line marks the expected HVC
    position. Dotted lines show the $\pm 2\sigma$ velocity range derived from the \HI\ spectrum. See
    text for details.}
  \label{fig: S441_ismfit}
\end{figure}

Since the stellar \TiII\ line only impinges on the wing of the expected HVC \CaK\ absorption, and is
very well fit by a single gaussian component, it does not diminish our ability to detect any
putative HVC absorption line. Figure~\ref{fig: S441_ismfit} shows the result of this fit.  We
include the strong stellar \CaK\ line, and a very weak unidentified line at rest wavelength
3932.25\AA\ ($\vLSR = -106\kms$). Both these latter features are well outside the region of interest
for HVC absorption. The lower panel shows the residual of the fit. The expected HVC position is
marked by the solid line, while the dotted lines mark the $\pm2\sigma$ line width limits (where
$\sigma$ is the gaussian width of the \HI\ spectrum).  We emphasize that only the high quality of
our Keck data allows us to cleanly remove the stellar absorption from the wing of the putative HVC
region, and place a significant lower distance limit on the cloud.

Since the S441 sightline is very close to the Mrk\,290 sightline, this significantly strengthens the
conclusions that can be drawn from the non-detection of both \CaK\ and \CaH, since we can be
confident of the level of HVC absorption expected. The worst-case expected HVC absorption toward
this line of sight is 38\mA\ (\CaK) and 19\mA\ (\CaH); both these would be clearly
visible. \S~\ref{sec: summary} contains further discussion on the interpretation of non-detections.

\begin{figure}[t]
  \epsscale{1.2}
  \plotone{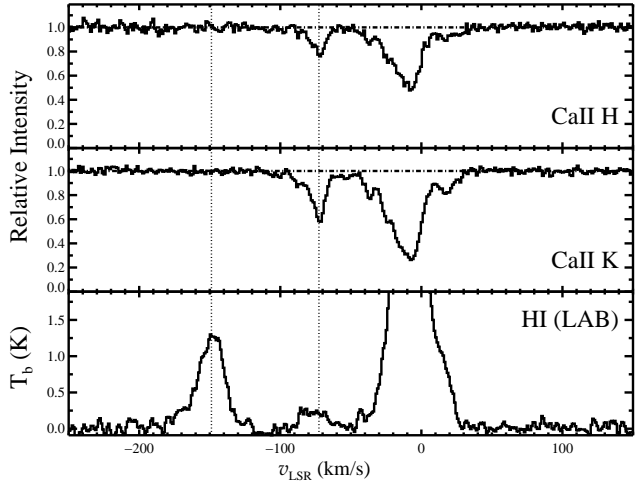}
  \caption{S135: No HVC absorption is evident at the expected velocity $\vLSR =
    -150\kms$. Multi-component IVC absorption can clearly be seen in both \CaHK\ absorption and
    \HI\ emission. The strongest absorption component is at $\vLSR = -77\kms$. Galactic absorption
    components, seen in both \CaK\ and \CaH, are visible between $-40 \simlt \vLSR \simlt 20\kms$,
    corresponding to the strong \HI\ emission.}
  \label{fig: S135}
\end{figure}

S135, $d \sim13\kpc$---The two stellar probes, S135 and S139 are both a few degrees from the edge
of \CC\ (See Figure~\ref{fig: ComplexC_aitoff}), at lower latitude than our earlier targets. For
S135, our most distant target, we see no indication of HVC absorption, as shown in Figure~\ref{fig:
  S135}. Unrelated IVC gas can clearly be seen in both \HI\ emission and absorption at $\vLSR =
-77\kms$ in both \CaHK. This IVC gas can be seen in the LAB data cube to connect smoothly with the
Galactic plane at lower latitudes. Although it is not shown in Figure~\ref{fig: S135}, we also see
this IVC in the \NaDb\ line, which is the stronger of the two \NaD\ doublet lines. The lack of HVC
absorption sets a lower distance limit of $12.8\pm3.2\kpc$.

\begin{figure}[t]
  \epsscale{1.2}
  \plotone{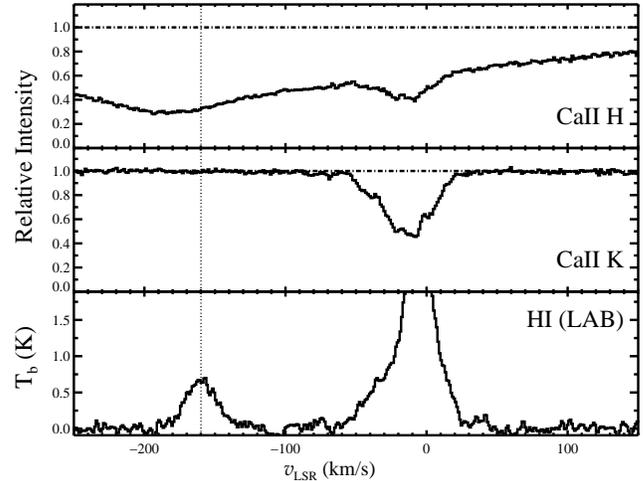}
  \caption{S139: No \CaK\ absorption can be seen corresponding to the HVC emission at $\vLSR =
    -160\kms$. The \CaH\ line falls in the core of the stellar \He\ line.}
  \label{fig: S139}
\end{figure}

S139,$d \sim8\kpc$---The optical spectra of S139 are our highest signal-to-noise
data. Figure~\ref{fig: S139} shows that there is clearly no absorption detected corresponding to the
strong HVC emission. The \CaH\ position is at the very core of the \He\ balmer line, but the \CaK\
data set a firm lower distance limit if $7.8\pm2.0\kpc$. The Galactic emission shows a broad
emission wing to $\vLSR \sim-50\kms$; \CaK\ absorption components are also seen in this range.
There is some evidence of IVC emission, with corresponding very weak absorption, in the range $\vLSR
= 70-100\kms$, but better data are needed to confirm this. The lack of HVC absorption sets a lower
distance limit.

\subsection{Upper Distance Limits}

Table~\ref{tab: upper_limits} gives the results of the two detections of \CC\ in absorption toward
background stars.  In this table, column (1) lists our target name. Stellar distances and measured
\HI\ column densities are given in Columns (2) and (3) respectively. Column (4) gives the measured
\CaK\ equivalent width and its $1\sigma$ error. Column (5) lists the corresponding column density
for this equivalent width, \NCa, assuming the gas is optically thin. Columns (3) and (5) are
combined to give the ratio \NCaNHI, which is sometimes called the ``abundance'' and denoted
\ACa.  This is given in column (6). Both sightlines have the same \CaII\ abundance to within
errors, and are also in good agreement with the values observed toward the QSO sightlines Mrk\,290
and PG\,1351+640, up to 20 degrees away (see below).  Such good agreement suggests that there are
not large variations of the \CaII\ abundance across the cloud.

\begin{deluxetable*}{lrrrrr}
\tabletypesize{\scriptsize}
\tablecaption{\label{tab: upper_limits}Summary of upper limits}
\tablewidth{0pt}
\tablecolumns{6}
\tablehead{
\colhead{Target} &
\colhead{Dist} &
\colhead{\NHI} &
\colhead{$\mbox{W}_{\lambda} \pm \sigEW$} &
\colhead{\NCa} &
\colhead{\ACa} \\
\colhead{} &
\colhead{(kpc)} &
\colhead{($\times 10^{19}\cm$)} &
\colhead{(\mA)} &
\colhead{($\times 10^{19}\cm$)} &
\colhead{}
}

\startdata
S437  & $11.3 \pm 2.8$ & $2.1 \pm 0.3$ & $46.9 \pm 2.0$ & $5.5 \pm 0.2$  & $26 \pm 4 \times 10^{-9}$ \\   
S674  & $10.4 \pm 2.6$ & $4.0 \pm 0.5$ & $76.5 \pm 1.8$ & $8.9 \pm 0.2$  & $22 \pm 3 \times 10^{-9}$ \\
\enddata

\tablecomments{Summary of measurements for upper limits. \NHI\ is taken from the LAB survey for both
  stars. The measured equivalent widths are for the \CaK\ HVC line. Column densities assume a linear
  curve-of-growth. The final column gives the \CaII\ to \HI\ column density ratio, which is referred
  to as ``abundance'' \ACa, and is measured in parts-per-billion.}
\end{deluxetable*}

\subsection{Lower Distance Limits}

Non-detections are harder to interpret than detections; one must be convinced that the lack of
absorption is not a random confluence of bad luck.  In general, one predicts the strength of the
absorption that should be seen if the star is behind the gas by using the measured \NCaNHI\ ratio in
the cloud, and the \HI\ column density toward the star to infer \NCa. This, in turn, provides the
expected equivalent width of the \CaII\ absorption line. We then attempt to show that the data are
of sufficient quality as to clearly detect absorption lines of that strength.

\citet{wakker-01-distance-metallicity} has argued that one should adopt a worst-case scenario to
interpret non-detections; in this view---which we adopt here---it is not sufficient that the
expected absorption be significantly stronger than the detection level, but we must account for the
case in which every factor works against a detection. For example, the gas-phase abundance may be
lower than expected, the column density may be lower than expected, and so on. This is quantified in
the notion of multiplicative ``safety factors'', which are given for a range of effects.  The
predicted absorption strength is reduced by all these safety factors to obtain a worst-case
absorption strength; only cases in which the noise level is less than this worst-case are deemed
``strong'' or significant non-detections. We refer the interested reader to the Appendix of
\citet{wakker-01-distance-metallicity}, for a more complete discussion of all these effects.

The comparison of large-beam 21\,cm measurements of \NHI\ to pencil-beam optical or UV data is
difficult because unresolved small-scale structure can potentially introduce systematic errors.  Our
\HI\ column densities are mostly taken from the LAB survey, which samples 36\arcmin\ of sky, or in
one case Effelsberg (9\arcmin).  Both of these sources have significantly larger beams than the
pencil beam of the optical data. This may be clearly seen, for example, in the spectra of S674
(Figure~\ref{fig: S674}), where the optical data show clear multi-component absorption structure,
yet the \HI\ emission spectrum does not. Systematic uncertainties are therefore likely to be present
since the radio and optical spectra sample different areas on the sky.

No comprehensive study exists comparing \NHI\ measured at a range of spatial scales toward a large
number of sightlines. The best study is that of \citet{wakker-etal-01-HI-spectra}, who compared
measurements at 36\arcmin, 9\arcmin and pencil-beam UV \Lya\ measurements on 6 QSO and 2 stellar
sightlines (only 2 toward \CC).  They concluded that most measurements at 9\arcmin\ are accurate to
within about 25\%, while the larger 36\arcmin\ beam gives an uncertainty of up to a factor of
3. This is in line with the results of \citet{savage-etal-00} who compared 10 UV \Lya\ measurements
of \NHI\ to those of the 21\arcmin\ beam of the NRAO 43m dish, finding that
$N(\HI)_{\Lya}/N(HI)_{21\,\mbox{cm}}$ ranges between 0.62 to 0.91. Thus we adopt a safety factor of
$2\times$ for the S441, where 9\arcmin\ Effelsberg data are available. For the other sightlines, a
safety factor of $3\times$ is warranted. We caution that variations may be still greater in cloud
cores \citep{wakker-oosterloo-putman-02-fine-structure}, and also note that both the above cited
works have only $8-10$ \Lya\ measurements of \NHI. Clearly this is the most uncertain aspect of our
results, and we are pursuing interferometer data (e.g.\ with the Allen Telescope Array [ATA]) for
our sightlines.

\begin{figure*}
  \epsscale{1.18}
  \plotone{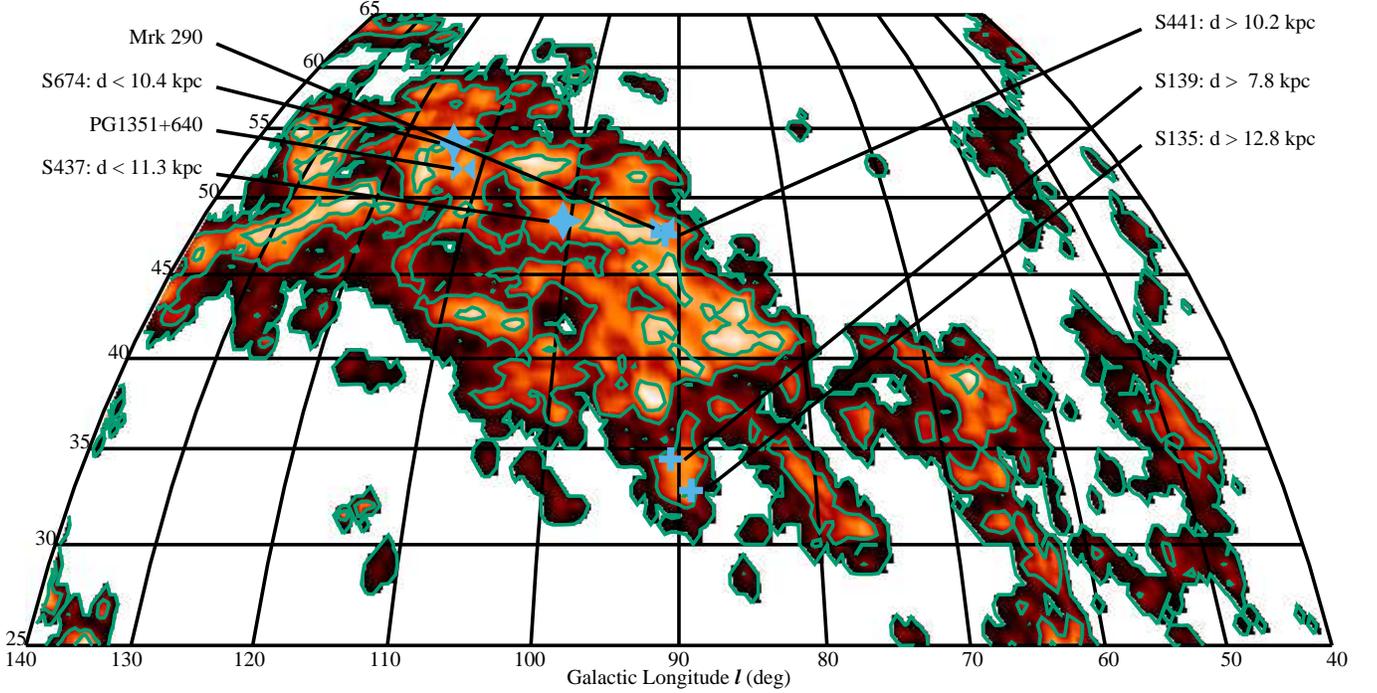}
  \caption{Location of SDSS stellar targets with respect to \HI\ emission of Complex~C. Contours are
    drawn at \logNHI\ = 18.2, 19.0, 19.7.}
  \label{fig: ComplexC_aitoff}
\end{figure*}

\CC\ shows little evidence of dust depletion \citep{collins-etal-03}, so we do not take into account
depletion effects. In the ideal case, the gas-phase abundance toward the stellar sightline will be
known, from measurements of nearby extra-galactic sightlines. This is the case for S441, which is
very close ($<0.5\degree$) to the Mrk\,290 sightline. For the S135 and S139 sightlines, however, we
must take into account possible variations in gas metallicity and ionization conditions. Since
\CaII\ is not the dominant ionization stage of Ca in the ISM and HVCs, a safety factor of $2\times$
is included to account for this \citep{wakker-01-distance-metallicity}.  Several authors have shown
evidence that the metallicity of \CC\ varies from QSO sightline to sightline in the range $0.1 -
0.3\Zsun$, independent of ionization effects \citep[e.g.][]{gibson-etal-01-ComplexC,
  collins-etal-07-ComplexC}. This may be a result of \CC\ mixing with local, enriched gas
\citep{gibson-etal-01-ComplexC, tripp-etal-03-ComplexC}. \CaII\ abundances have been measured toward
only two sightlines: Mrk\,290 [$A(\CaII) = 21 \times 10^{-9}$] and PG\,1351+640 [$A(\CaII) = 18
\times 10^{-9}$] \citep[][]{wakker-etal-96-ComplexC-CaII-abundance,
  wakker-01-distance-metallicity}. Mrk\,290 is closest to the S135 and S139 sightlines, so we adopt
the measured abundance, and include a safety factor of $3\times$, in line with the metallicity
variations.

Table~\ref{tab: lower_limits} provides a summary of the process used to determine the significance
of the non-detections.  Column (1) provides target name. Column (2) gives the stellar distance. The
measured \NHI\ toward the star is listed in column (3). Column (4) shows the predicted absorption
equivalent width, $\mbox{W}_{\lambda, pred}$. To calculate this value, we assume the gas is
optically thin, and adopt $\ACa = 21\times10^{-9}$. Column (5) lists the safety factor adopted,
with column (6) giving the resulting strength of any putative absorption line, reduced by this
safety factor. Column (7) lists the $1\sigma$ equivalent width error, integrating over a line-width
determined from the \HI\ data. Finally, column (8) gives the significance of the
non-detection. Following \citet{wakker-01-distance-metallicity}, a significance greater than 1 is
considered a ``strong'' lower distance limit, with values lower than 1 providing ``weak''
limits. 

Table~\ref{tab: lower_limits} shows that even under pessimistic assumptions, the excellent quality
of the Keck data gives us confidence in our non-detections. For the sightline toward S441, the
safety factor according to the above prescription is significantly smaller than the other two
sightlines, since the \HI\ column density is more accurately known, and the ${\rm Ca}^+$ abundance
of the gas in this region is well measured. Nevertheless, a factor of $2\times$ is almost certainly
too optimistic. In line with this concern, we include in Table~\ref{tab: lower_limits} a second
assessment of the significance of this non-detection, using the same $18\times$ factor as for S135
and S139. Even under this scenario, any putative HVC absorption would still be detected. We conclude
that all our non-detections provide strong lower distance limits to \CC.

\begin{deluxetable*}{lrrrrrrrr}
\tabletypesize{\scriptsize}
\tablecaption{\label{tab: lower_limits}Summary of lower limits}
\tablewidth{0pt}
\tablecolumns{9}
\tablehead{
\colhead{Target} &
\colhead{Dist} &
\colhead{\NHI} &
\colhead{$\mbox{W}_{\lambda, pred}$} &
\colhead{Saf} &
\colhead{$\mbox{W}_{\lambda, saf}$} &
\colhead{\sigEW} &
\colhead{Sig} \\
\colhead{} &
\colhead{(kpc)} &
\colhead{($\times 10^{19}\cm$)} &
\colhead{(\mA)} &
\colhead{} &
\colhead{(\mA)} &
\colhead{(\mA)} &
\colhead{}
}

\startdata
S135  & $12.8 \pm 3.2$ & $4.9 \pm 0.4$ & 88.8  &  18    &  4.9    &  $\pm 1.6$      &  3.1 \\
S139  & $ 7.8 \pm 2.0$ & $3.3 \pm 0.4$ & 60.2  &  18    &  3.3    &  $\pm 1.1$      &  3.0 \\
S441  & $10.2 \pm 2.6$ & $4.2 \pm 0.2$ & 75.3  &   2    & 37.7    &  $\pm 1.7$      & 22.2 \\
S441  & $10.2 \pm 2.6$ & $4.2 \pm 0.2$ & 75.3  &  18    &  4.2    &  $\pm 1.7$      &  2.5 \\
\enddata

\tablecomments{Summary of measurements for lower limits. The source of distances and \NHI\ is
  described in the text. We give the predicted \CaK\ equivalent width before and after taking into
  account the safety factors, as well as the safety factor adopted for each target. These
  contributions to these numbers are discussed in \S~\ref{sec: summary}. The significance of the
  non-detections is given as the ratio of $\mbox{W}_{\lambda, saf} / \sigEW$. The final column
  tabulates the abundance \NCaNHI\ used to derived the predicted equivalent widths, also discussed
  in \S~\ref{sec: summary}.}
\end{deluxetable*}

\subsection{Summary of Results} 
\label{sec: summary}

We summarize our results in Table~\ref{tab: summary}. This table gives the Galactic co-ordinates and
stellar distances for each target, and indicates whether the target provides and upper (U) or lower
(L) distance limit to \CC. Figure~\ref{fig: ComplexC_aitoff} shows an \HI\ column density map of
\CC\ in Galactic co-ordinates, using an orthogonal projection. Contours are drawn at $\logNHI\ =
18.2, 19.0, 19.7$.  The stellar lines of sight are marked with star symbols in the case of
detections, and crosses in the case of non-detections. The stars are labeled, along with the
corresponding stellar distances. We also mark the position of the Mrk\,290 and PG\,1351+640
sightlines (hourglasses), along which the \CaII\ abundance is measured.

\begin{deluxetable}{lrrrr}
\tabletypesize{\scriptsize}
\tablecaption{\label{tab: summary}Summary of distance limits}
\tablewidth{0pt}
\tablecolumns{5}
\tablehead{
\colhead{Target} &
\colhead{$l$} &
\colhead{$b$} &
\colhead{Dist} &
\colhead{Type} \\
\colhead{} &
\colhead{(deg)} &
\colhead{(deg)} &
\colhead{(kpc)} &
\colhead{}
}
\startdata
S135 &   89.1 & 32.8 & $12.8 \pm 3.2$ & L \\
S139 &   90.6 & 34.4 & $ 7.8 \pm 2.0$ & L \\
S437 &  100.7 & 48.4 & $11.3 \pm 2.8$ & U \\
S441 &   91.2 & 47.5 & $10.2 \pm 2.6$ & L \\
S674 &  114.0 & 54.0 & $10.4 \pm 2.6$ & U \\
\enddata

\tablecomments{Summary of our distance limits. Co-ordinates are galactic and are provided to the
  nearest $1/10^{th}$ of a degree. The distance to each star is listed, with an indication whether
  the star provides a lower (L), or upper (U) distance limit.}
\end{deluxetable}

\begin{figure*}
  \epsscale{1.0}
  \plotone{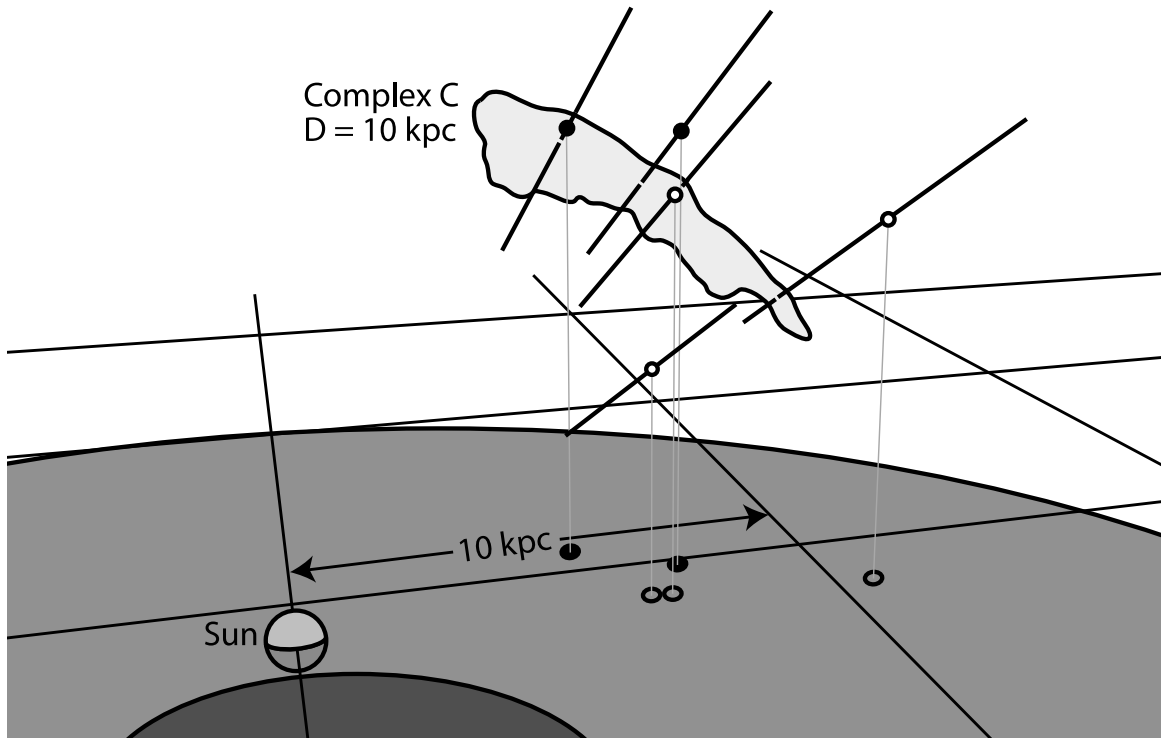}
  \caption{Diagram showing the position of \CC\ in relation to the Sun. The inner Galaxy is
    indicated by the dark region at the bottom. HVC detection are marked by full circles;
    non-detection by open circles. The position of stellar targets projected onto the Galactic plane
    are also shown. $1\,\sigma$ distance error bars are also shown. Note that the shape of \CC\ here
    is {\it not} an accurate depiction, and is indicative only.}
  \label{fig: ComplexC_diagram}
\end{figure*}

For orientation, we also show \CC\ in perspective view in Figure~\ref{fig: ComplexC_diagram}. This
figure shows the position of the gas in relation to the Galactic centre and the Solar
position. Solid circles in this diagram are the HVC detections, S437 and S674. Open circles show the
positions of the non-detections---S135, S139 and S441. For all stars, we also show their position
projected onto the plane of the Galaxy. $1\,\sigma$ distance error bars are indicated for each
star. The shape and orientation of \CC\ in this diagram should be taken as indicative
only\footnote{Indeed, recent \HI\ data show \CC\ extends much closer to the plane, although it is no
  longer a single coherent structure at these latitudes.}. An accurate representation would require
more distance constraints, especially in the low-latitude, low-longitude regions (right-most part in
this representation).

Taken together, our distance limits imply a canonical distance of $10 \pm 2.5\kpc$ to \CC\ in the
higher latitude regions. Toward S674 at the highest longitudes, the HVC is constrained to be closer
than $10.4 \pm 2.6\kpc$. In the middle latitudes, both the upper limit from S437 ($d <
11.3\pm2.8\kpc$) and lower set by S441 ($d > 10.2\pm2.6\kpc$) are consistent with the 10\kpc\
distance.  At lower latitudes, we have only lower limits from S135 ($d > 12.8\pm3.2\kpc$) and S139
($d > 7.8 \pm 2.0\kpc$). While both these stars are consistent with a 10\kpc\ distance for \CC, we
have no limits for the low latitude parts of the complex, and cannot exclude the possibility of a
distance gradient. It would not be surprising if the lower latitude, lower longitude parts of the
complex were more distant given the large angular size of \CC. Observations at high latitudes of
closer stars in the range $6-8\kpc$, and more distant stars in the lower longitude, lower latitude
regions, are needed to substantiate this speculation.  We stress that the $\sim25\%$ distance
accuracy is the current limit of stellar classification techniques for horizontal-branch stars, and
applies to {\it all} results; a systematic calibration effort is required to reduce this
uncertainty.

\section{Discussion}

For ease of discussion, we first consider \CC\ to be at a uniform distance of 10\kpc. With such a
large projected size, this is unlikely unless \CC\ has a curved geometry.  Using the total flux for
\CC\ \citep{wakker-vanwoerden-91-catalogue}, our distance limit implies a mass for \CC\ of \CCMHI.
To calculate the total mass, we include a factor of 1.4 to account for helium, and an ionization
fraction $M_{\rm H+} = 0.18\,\MHI$ \citep{wakker-etal-99-nature, sembach-etal-03-OVI-HVC}. Since
there is no indication of molecular gas \citep{murphy-etal-00-ComplexC-FUSE,
  richter-etal-01-MolecularGas}, we take $M_{\rm H} = M_{\rm H+} + \MHI$, and derive a total mass
for \CC\ of \CCMtot.

For comparison, \CC\ has an \HI\ mass of order the mass that \citet{lockman-03-ComplexH} derived for
Complex~H by assuming that it is a satellite of the MW merging with the outer disk.  Of the recently
discovered low-luminosity dwarf galaxies recently discovered in the SDSS
\citep[e.g.][]{willman-etal-05-willmanI, willman-etal-05-UMa, belokurov-etal-06-bootes,
  belokurov-etal-07-MW-dwarfs}, only Leo~T has been shown to have associated neutral gas
\citep{ryan-weber-etal-07-LeoT-HI}; the \HI\ mass of \CC\ is more than an order of magnitude more
than that of Leo~T, and is comparable to other local group dwarf irregulars, such as Pegasus,
DDO\,210 and LGS\,3 \citep{mateo-98}. \CC\ is also an order of magnitude more massive than the HVCs
surrounding the M31/M33 system \citep{westmeier-etal-05-M31}. Despite the large masses of neutral
gas, a variety of searches have failed to find any evidence for an associated stellar content with
any HVC \citep[e.g.][]{willman-etal-02-stellar-content, simon-blitz-02-CHVC-stellar,
  siegel-etal-05-stellar-search, simon-etal-06-ComplexH}, and there is no evidence of a connection
between \CC\ and any dwarf galaxies.

The calculation of the mass flux onto the Galaxy that \CC\ provides is hampered by our ignorance of
the tangential velocity, which limits our ability to determine the vertical velocity with respect to
the disk. We attempt to average over much of our ignorance by considering the average mass flow over
the whole accretion timescale (i.e. the time it takes for the whole complex to accrete).  The
furthest gas from the plane, at highest latitudes is that between $b = 55-60\degree$---we choose a
representative direction (l, b) = ($120\degree, 58\degree$). At $d=10\kpc$, this gas is $\sim8\kpc$ above
the disk, and has a line-of-sight velocity $\vLSR = -126\kms$. To remove the disk rotation, we
convert to ``deviation velocity'', which is the amount by which the gas deviates from a model of
Galactic rotation \citep{deheij-etal-02-catalogue}, giving $\vdev = -105\kms$. We consider three
cases: first, we assume that the tangential component of the velocity is equal to the radial
component, and consider both extremes, in which this tangential component is directed toward and
away from the disk. This results in a range of vertical velocities $v_z = -34 - -145\kms$, where the
negative indicates a direction toward the disk. The case in which the motion is purely radial
(i.e. no tangential component) is intermediate to these two extremes, having simply $v_z =
-105\,sin(58\degree) = -89\kms$. This range of velocities results in a range of accretion timescales
$60 - 250\Myr$, and a corresponding mass flux in the range $0.03 - 0.14\Msunyr$. It is worth noting
that \citet[][]{tripp-etal-03-ComplexC} have argued that the lower latitude parts of \CC\ show signs
of interaction with the thick disk or lower halo.

The condensing cloud model \citep{sommer-larsen-06, peek-etal-07-accretion} predicts
$\sim0.2\Msunyr$ of accretion coming from HVCs. Meanwhile, chemical evolution models require such
infall to reproduce the observed metallicity distribution in the disk
\citep{alibes-etal-01-infall-rate, fenner-gibson-03}. The most recent calculations suggest an infall
rate of $\sim1\Msunyr$ about 5\Gyr\ ago, falling roughly to half that at the present epoch
\citep{chiappini-etal-01}. Thus \CC\ may provide a substantial amount of the required mass flux on
the Galaxy, although the question of how this \HI\ is converted into stars remains.

Now that we have established the distance to \CC, we can also attempt to put some constraints on
physical parameters, such as length scales and density. Clearly the density will vary from sightline
to sightline for a large, inhomogeneous structure like \CC, and the numbers we derive here should be
taken as indicative only. For a canonical distance $d = 10\kpc$, simple geometry implies a
transverse scale factor $\sim0.175\kpc\,\rm{deg}^{-1}$. If we take two representative points on the
extreme edges of the cloud $(l, b) = (90, 50), (110, 40)$, the maximum angular extent across \CC\ is
$\theta \approx 17\degree$, which corresponds to a physical distance $L \approx 3.0\kpc$. To assess
the length of \CC\ we take the points $(l, b) = (30, 15), (130, 55)$\footnote{Note that
  Figure~\ref{fig: ComplexC_aitoff} does not show the full extent of \CC.}, which correspond to a
length $\sim15\kpc$. Lower latitude and longitude gas is present, but it's direct connection with
\CC\ is ambiguous, since it blends with disk emission.  By comparison, the Magellanic Stream is
approximately $10 \times 100\kpc$ in size, with a mass of $\sim2\times 10^8\Msun$ for an assumed
distance of 55\kpc \citep{putman-etal-03-mshvc}. If we further assume that the complex is as deep as
it is wide, then the average density, $\langle n \rangle$, along the line of sight may be
computed. Towards the two stars which provide upper limits, S437 and S674, we then have $log\,n =
-2.6$ and $-2.4$ respectively.

\section{Conclusion}

We have used the presence and absence of \CaK\ absorption in 5 stars aligned with the high-velocity
cloud \CC\ to derive a distance to the complex of $10\pm2.5\kpc$.  At high latitude, we detect HVC
gas in absorption towards stars $\sim10$ and $\sim11\kpc$ away, setting upper distance limits. At
similar latitudes, a non-detection provides a lower limit of $\sim10\kpc$. Non-detections in stars
at $\sim8$ and $\sim13\kpc$ at lower latitudes are consistent with this distance. Since the stellar
distances are accurate to $\sim25\%$, a canonical distance of $10\pm2.5\kpc$ is set, but we cannot
exclude the possibility of a distance gradient, which would mean larger distances for lower latitude
parts of \CC.  Indeed, such a distance gradient would be expected when one considers the large
angular size of the complex.

The distance implies an \HI\ mass for \CC\ of \CCMHI. Applying corrections for helium and ionized
gas yields total mass of \CCMtot. We derive a mass inflow rate in the range $0.03 - 0.14\Msunyr$,
with the uncertainty mostly coming from the unknown tangential velocity of \CC. If there is no
tangential velocity component, this inflow rate is $\sim0.1\Msunyr$. Thus \CC\ may provide a
significant fraction of the infall rate predicted by the condensing cloud model. At the measured
distance, \CC\ is some $\sim3\kpc$ across, with an average density of order $log\,n = -2.5$.  Our
accurate distance contributes to a picture in which the large HVC complexes are nearby Galactic
objects, with many of them now known to be within $d < 10-15\kpc$ \citep{vanwoerden-etal-99-nature,
  thom-etal-06-complexWB, wakker-etal-07-ComplexC, wakker-etal-07-I-distances}. The question of
whether the compact HVCs are part of this same population remains an open question, although
condensation models place these small, isolated clouds at larger distances. With the first elements
to a solution to the high-velocity cloud distance problem now well established, continuing efforts
will provide a more complete census of neutral gas in the Milky Way halo.

\acknowledgements 

We thank Tobias Westmeier for providing the HVC data from which Figure~\ref{fig: ComplexC_aitoff}
was made, and for the Effelsberg spectrum towards S441.  The optical data presented herein were
obtained at the W.~M.~Keck Observatory, which is operated as a scientific partnership among the
California Institute of Technology, the University of California and the National Aeronautics and
Space Administration. The Observatory was made possible by the generous financial support of the
W.~M.~Keck Foundation. The authors wish to recognize and acknowledge the very significant cultural
role and reverence that the summit of Mauna Kea has always had within the indigenous Hawaiian
community.  We are most fortunate to have the opportunity to conduct observations from this
mountain.

This work has made use of the NIST Atomic Spectra Database (v3.1), available on-line at
http://physics.nist.gov/asd3. C.~T. acknowledges partial support from NASA through the American
Astronomical Society's Small Research Grant Program, and NASA grant NNG06GC36G. C.~H. acknowledges
support from NSF grant AST-0406987.

\bibliography{MASTER}

\begin{thebibliography}{65}
\expandafter\ifx\csname natexlab\endcsname\relax\def\natexlab#1{#1}\fi

\bibitem[{{Alib{\'e}s} {et~al.}(2001){Alib{\'e}s}, {Labay}, \&
  {Canal}}]{alibes-etal-01-infall-rate}
{Alib{\'e}s}, A., {Labay}, J., \& {Canal}, R. 2001, \aap, 370, 1103

\bibitem[{{Belokurov} {et~al.}(2007){Belokurov}, {Zucker}, {Evans}, {Kleyna},
  {Koposov}, {Hodgkin}, {Irwin}, {Gilmore}, {Wilkinson}, {Fellhauer},
  {Bramich}, {Hewett}, {Vidrih}, {De Jong}, {Smith}, {Rix}, {Bell}, {Wyse},
  {Newberg}, {Mayeur}, {Yanny}, {Rockosi}, {Gnedin}, {Schneider}, {Beers},
  {Barentine}, {Brewington}, {Brinkmann}, {Harvanek}, {Kleinman}, {Krzesinski},
  {Long}, {Nitta}, \& {Snedden}}]{belokurov-etal-07-MW-dwarfs}
{Belokurov}, V., {Zucker}, D.~B., {Evans}, N.~W., {Kleyna}, J.~T., {Koposov},
  S., {Hodgkin}, S.~T., {Irwin}, M.~J., {Gilmore}, G., {Wilkinson}, M.~I.,
  {Fellhauer}, M., {Bramich}, D.~M., {Hewett}, P.~C., {Vidrih}, S., {De Jong},
  J.~T.~A., {Smith}, J.~A., {Rix}, H.-W., {Bell}, E.~F., {Wyse}, R.~F.~G.,
  {Newberg}, H.~J., {Mayeur}, P.~A., {Yanny}, B., {Rockosi}, C.~M., {Gnedin},
  O.~Y., {Schneider}, D.~P., {Beers}, T.~C., {Barentine}, J.~C., {Brewington},
  H., {Brinkmann}, J., {Harvanek}, M., {Kleinman}, S.~J., {Krzesinski}, J.,
  {Long}, D., {Nitta}, A., \& {Snedden}, S.~A. 2007, \apj, 654, 897

\bibitem[{{Belokurov} {et~al.}(2006){Belokurov}, {Zucker}, {Evans},
  {Wilkinson}, {Irwin}, {Hodgkin}, {Bramich}, {Irwin}, {Gilmore}, {Willman},
  {Vidrih}, {Newberg}, {Wyse}, {Fellhauer}, {Hewett}, {Cole}, {Bell}, {Beers},
  {Rockosi}, {Yanny}, {Grebel}, {Schneider}, {Lupton}, {Barentine},
  {Brewington}, {Brinkmann}, {Harvanek}, {Kleinman}, {Krzesinski}, {Long},
  {Nitta}, {Smith}, \& {Snedden}}]{belokurov-etal-06-bootes}
{Belokurov}, V., {Zucker}, D.~B., {Evans}, N.~W., {Wilkinson}, M.~I., {Irwin},
  M.~J., {Hodgkin}, S., {Bramich}, D.~M., {Irwin}, J.~M., {Gilmore}, G.,
  {Willman}, B., {Vidrih}, S., {Newberg}, H.~J., {Wyse}, R.~F.~G., {Fellhauer},
  M., {Hewett}, P.~C., {Cole}, N., {Bell}, E.~F., {Beers}, T.~C., {Rockosi},
  C.~M., {Yanny}, B., {Grebel}, E.~K., {Schneider}, D.~P., {Lupton}, R.,
  {Barentine}, J.~C., {Brewington}, H., {Brinkmann}, J., {Harvanek}, M.,
  {Kleinman}, S.~J., {Krzesinski}, J., {Long}, D., {Nitta}, A., {Smith}, J.~A.,
  \& {Snedden}, S.~A. 2006, \apjl, 647, L111

\bibitem[{{Bregman}(1980)}]{bregman-80}
{Bregman}, J.~N. 1980, \apj, 236, 577

\bibitem[{{Chiappini} {et~al.}(2001){Chiappini}, {Matteucci}, \&
  {Romano}}]{chiappini-etal-01}
{Chiappini}, C., {Matteucci}, F., \& {Romano}, D. 2001, \apj, 554, 1044

\bibitem[{{Collins} {et~al.}(2003){Collins}, {Shull}, \&
  {Giroux}}]{collins-etal-03}
{Collins}, J.~A., {Shull}, J.~M., \& {Giroux}, M.~L. 2003, \apj, 585, 336

\bibitem[{{Collins} {et~al.}(2007){Collins}, {Shull}, \&
  {Giroux}}]{collins-etal-07-ComplexC}
---. 2007, \apj, 657, 271

\bibitem[{{Danly} {et~al.}(1993){Danly}, {Albert}, \&
  {Kuntz}}]{danly-etal-93-ComplexM}
{Danly}, L., {Albert}, C.~E., \& {Kuntz}, K.~D. 1993, \apjl, 416, L29+

\bibitem[{{de Avillez}(2000)}]{deAvillez-00-galfountain}
{de Avillez}, M.~A. 2000, \apss, 272, 23

\bibitem[{{de Heij} {et~al.}(2002){de Heij}, {Braun}, \&
  {Burton}}]{deheij-etal-02-catalogue}
{de Heij}, V., {Braun}, R., \& {Burton}, W.~B. 2002, \aap, 391, 159

\bibitem[{{Dorman}(1992)}]{dorman-92-HB-isochrones}
{Dorman}, B. 1992, \apjs, 81, 221

\bibitem[{{Fenner} \& {Gibson}(2003)}]{fenner-gibson-03}
{Fenner}, Y. \& {Gibson}, B.~K. 2003, Publ. Astr. Soc. Aus., 20, 189

\bibitem[{{Fox} {et~al.}(2004){Fox}, {Savage}, {Wakker}, {Richter}, {Sembach},
  \& {Tripp}}]{fox-etal-04-ComplexC}
{Fox}, A.~J., {Savage}, B.~D., {Wakker}, B.~P., {Richter}, P., {Sembach},
  K.~R., \& {Tripp}, T.~M. 2004, \apj, 602, 738

\bibitem[{{Gibson} {et~al.}(2001){Gibson}, {Giroux}, {Penton}, {Stocke},
  {Shull}, \& {Tumlinson}}]{gibson-etal-01-ComplexC}
{Gibson}, B.~K., {Giroux}, M.~L., {Penton}, S.~V., {Stocke}, J.~T., {Shull},
  J.~M., \& {Tumlinson}, J. 2001, \aj, 122, 3280

\bibitem[{{Girardi} {et~al.}(2004){Girardi}, {Grebel}, {Odenkirchen}, \&
  {Chiosi}}]{girardi-etal-04-isochrones}
{Girardi}, L., {Grebel}, E.~K., {Odenkirchen}, M., \& {Chiosi}, C. 2004, \aap,
  422, 205

\bibitem[{{Haffner} {et~al.}(2003){Haffner}, {Reynolds}, {Tufte}, {Madsen},
  {Jaehnig}, \& {Percival}}]{haffner-etal-03-WHAM-survey}
{Haffner}, L.~M., {Reynolds}, R.~J., {Tufte}, S.~L., {Madsen}, G.~J.,
  {Jaehnig}, K.~P., \& {Percival}, J.~W. 2003, \apjs, 149, 405

\bibitem[{{Houck} \& {Bregman}(1990)}]{houck-bregman-90}
{Houck}, J.~C. \& {Bregman}, J.~N. 1990, \apj, 352, 506

\bibitem[{{Hulsbosch} \& {Raimond}(1966)}]{hulsbosch-raimond-66-survey}
{Hulsbosch}, A.~N.~M. \& {Raimond}, E. 1966, \bain, 18, 413

\bibitem[{{Kalberla} {et~al.}(2005){Kalberla}, {Burton}, {Hartmann}, {Arnal},
  {Bajaja}, {Morras}, \& {P{\"o}ppel}}]{kalberla-etal-05-LAB}
{Kalberla}, P.~M.~W., {Burton}, W.~B., {Hartmann}, D., {Arnal}, E.~M.,
  {Bajaja}, E., {Morras}, R., \& {P{\"o}ppel}, W.~G.~L. 2005, \aap, 440, 775

\bibitem[{{Kaufmann} {et~al.}(2006){Kaufmann}, {Mayer}, {Wadsley}, {Stadel}, \&
  {Moore}}]{kaufmann-etal-06-condensation-simulation}
{Kaufmann}, T., {Mayer}, L., {Wadsley}, J., {Stadel}, J., \& {Moore}, B. 2006,
  \mnras, 370, 1612

\bibitem[{Lee {et~al.}(2008)Lee, Beers, Sivarani, Johnson, An, Wilhelm,
  Allende-Prieto, Fiorentin, Bailer-Jones, Norris, Yanny, Rockosi, Newberg,
  Cudworth, \& Pan}]{lee-etal-08-sspp-I}
Lee, Y.-S., Beers, T.~C., Sivarani, T., Johnson, J., An, D., Wilhelm, R.,
  Allende-Prieto, C., Fiorentin, P.~R., Bailer-Jones, C.~A., Norris, J.~E.,
  Yanny, B., Rockosi, C.~M., Newberg, H.~J., Cudworth, K.~M., \& Pan, K. 2008,
  \apj, submitted

\bibitem[{{Lockman}(2003)}]{lockman-03-ComplexH}
{Lockman}, F.~J. 2003, \apjl, 591, L33

\bibitem[{{Maller} \& {Bullock}(2004)}]{maller-bullock-04}
{Maller}, A.~H. \& {Bullock}, J.~S. 2004, \mnras, 355, 694

\bibitem[{{Mateo}(1998)}]{mateo-98}
{Mateo}, M.~L. 1998, \araa, 36, 435

\bibitem[{{Mathewson} {et~al.}(1974){Mathewson}, {Cleary}, \&
  {Murray}}]{mathewson-etal-74-MS}
{Mathewson}, D.~S., {Cleary}, M.~N., \& {Murray}, J.~D. 1974, \apj, 190, 291

\bibitem[{{Meggers} {et~al.}(1976){Meggers}, {Corliss}, \&
  {Scribner}}]{meggers-etal-75-NIST-Ti}
{Meggers}, W.~F., {Corliss}, C.~H., \& {Scribner}, B.~F. 1976 (National Bureau
  of Standards Monograph 145, Washington: US Government Printing Office
  (USGPO), 1975)

\bibitem[{{Muller} {et~al.}(1963){Muller}, {Oort}, \&
  {Raimond}}]{muller-etal-63-HVC-discovery}
{Muller}, C.~A., {Oort}, J.~H., \& {Raimond}, E. 1963, Comptes Rendus de
  l'Acad{\'e}mie des Sciences Paris, 257, 1661

\bibitem[{{Murphy} {et~al.}(2000){Murphy}, {Sembach}, {Gibson}, {Shull},
  {Savage}, {Roth}, {Moos}, {Green}, {York}, \&
  {Wakker}}]{murphy-etal-00-ComplexC-FUSE}
{Murphy}, E.~M., {Sembach}, K.~R., {Gibson}, B.~K., {Shull}, J.~M., {Savage},
  B.~D., {Roth}, K.~C., {Moos}, H.~W., {Green}, J.~C., {York}, D.~G., \&
  {Wakker}, B.~P. 2000, \apjl, 538, L35

\bibitem[{{Oort}(1966)}]{oort-66-HVC-origins}
{Oort}, J.~H. 1966, \bain, 18, 421

\bibitem[{{Oosterloo} {et~al.}(2007){Oosterloo}, {Fraternali}, \&
  {Sancisi}}]{oosterloo-etal-07-NGC891}
{Oosterloo}, T., {Fraternali}, F., \& {Sancisi}, R. 2007, \aj, 134, 1019

\bibitem[{{Pagel} \& {Patchett}(1975)}]{pagel-patchett-75}
{Pagel}, B.~E.~J. \& {Patchett}, B.~E. 1975, \mnras, 172, 13

\bibitem[{{Peek} {et~al.}(2007){Peek}, {Putman}, \&
  {Sommer-Larsen}}]{peek-etal-07-accretion}
{Peek}, J.~E.~G., {Putman}, M.~E., \& {Sommer-Larsen}, J. 2007, ArXiv e-prints,
  705

\bibitem[{{Putman} {et~al.}(2003){Putman}, {Staveley-Smith}, {Freeman},
  {Gibson}, \& {Barnes}}]{putman-etal-03-mshvc}
{Putman}, M.~E., {Staveley-Smith}, L., {Freeman}, K.~C., {Gibson}, B.~K., \&
  {Barnes}, D.~G. 2003, \apj, 586, 170

\bibitem[{{Putman} {et~al.}(2004){Putman}, {Thom}, {Gibson}, \&
  {Staveley-Smith}}]{putman-etal-04-sgr}
{Putman}, M.~E., {Thom}, C., {Gibson}, B.~K., \& {Staveley-Smith}, L. 2004,
  \apjl, 603, L77

\bibitem[{{Richter} {et~al.}(2001){Richter}, {Sembach}, {Wakker}, \&
  {Savage}}]{richter-etal-01-MolecularGas}
{Richter}, P., {Sembach}, K.~R., {Wakker}, B.~P., \& {Savage}, B.~D. 2001,
  \apjl, 562, L181

\bibitem[{{Ryan-Weber} {et~al.}(2007){Ryan-Weber}, {Begum}, {Oosterloo}, {Pal},
  {Irwin}, {Belokurov}, {Evans}, \& {Zucker}}]{ryan-weber-etal-07-LeoT-HI}
{Ryan-Weber}, E.~V., {Begum}, A., {Oosterloo}, T., {Pal}, S., {Irwin}, M.~J.,
  {Belokurov}, V., {Evans}, N.~W., \& {Zucker}, D.~B. 2007, \mnras, in press
  (arXiv:0711.2979)

\bibitem[{{Ryans} {et~al.}(1997){Ryans}, {Keenan}, {Sembach}, \&
  {Davies}}]{ryans-etal-97-IVArch-ComplexM}
{Ryans}, R.~S.~I., {Keenan}, F.~P., {Sembach}, K.~R., \& {Davies}, R.~D. 1997,
  \mnras, 289, 83

\bibitem[{{Savage} {et~al.}(2000){Savage}, {Wakker}, {Jannuzi}, {Bahcall},
  {Bergeron}, {Boksenberg}, {Hartig}, {Kirhakos}, {Murphy}, {Sargent},
  {Schneider}, {Turnshek}, \& {Wolfe}}]{savage-etal-00}
{Savage}, B.~D., {Wakker}, B., {Jannuzi}, B.~T., {Bahcall}, J.~N., {Bergeron},
  J., {Boksenberg}, A., {Hartig}, G.~F., {Kirhakos}, S., {Murphy}, E.~M.,
  {Sargent}, W.~L.~W., {Schneider}, D.~P., {Turnshek}, D., \& {Wolfe}, A.~M.
  2000, \apjs, 129, 563

\bibitem[{{Schwarz} {et~al.}(1995){Schwarz}, {Wakker}, \& {van
  Woerden}}]{schwarz-etal-95}
{Schwarz}, U.~J., {Wakker}, B.~P., \& {van Woerden}, H. 1995, \aap, 302, 364

\bibitem[{{Sembach} \& {Savage}(1992)}]{sembach-savage-92-EW}
{Sembach}, K.~R. \& {Savage}, B.~D. 1992, \apjs, 83, 147

\bibitem[{{Sembach} {et~al.}(2003){Sembach}, {Wakker}, {Savage}, {Richter},
  {Meade}, {Shull}, {Jenkins}, {Sonneborn}, \&
  {Moos}}]{sembach-etal-03-OVI-HVC}
{Sembach}, K.~R., {Wakker}, B.~P., {Savage}, B.~D., {Richter}, P., {Meade}, M.,
  {Shull}, J.~M., {Jenkins}, E.~B., {Sonneborn}, G., \& {Moos}, H.~W. 2003,
  \apjs, 146, 165

\bibitem[{{Siegel} {et~al.}(2005){Siegel}, {Majewski}, {Gallart}, {Sohn},
  {Kunkel}, \& {Braun}}]{siegel-etal-05-stellar-search}
{Siegel}, M.~H., {Majewski}, S.~R., {Gallart}, C., {Sohn}, S.~T., {Kunkel},
  W.~E., \& {Braun}, R. 2005, \apj, 623, 181

\bibitem[{{Simon} \& {Blitz}(2002)}]{simon-blitz-02-CHVC-stellar}
{Simon}, J.~D. \& {Blitz}, L. 2002, \apj, 574, 726

\bibitem[{{Simon} {et~al.}(2006){Simon}, {Blitz}, {Cole}, {Weinberg}, \&
  {Cohen}}]{simon-etal-06-ComplexH}
{Simon}, J.~D., {Blitz}, L., {Cole}, A.~A., {Weinberg}, M.~D., \& {Cohen}, M.
  2006, \apj, 640, 270

\bibitem[{{Sirko} {et~al.}(2004){Sirko}, {Goodman}, {Knapp}, {Brinkmann},
  {Ivezi{\' c}}, {Knerr}, {Schlegel}, {Schneider}, \& {York}}]{sirko-etal-04a}
{Sirko}, E., {Goodman}, J., {Knapp}, G.~R., {Brinkmann}, J., {Ivezi{\' c}}, {\v
  Z}., {Knerr}, E.~J., {Schlegel}, D., {Schneider}, D.~P., \& {York}, D.~G.
  2004, \aj, 127, 899

\bibitem[{{Sommer-Larsen}(2006)}]{sommer-larsen-06}
{Sommer-Larsen}, J. 2006, \apjl, 644, L1

\bibitem[{{Thom}(2006)}]{thom-06-PhD-thesis}
{Thom}, C. 2006, PhD thesis, Swinburne University of Technology

\bibitem[{{Thom} {et~al.}(2006){Thom}, {Putman}, {Gibson}, {Christlieb},
  {Flynn}, {Beers}, {Wilhelm}, \& {Lee}}]{thom-etal-06-complexWB}
{Thom}, C., {Putman}, M.~E., {Gibson}, B.~K., {Christlieb}, N., {Flynn}, C.,
  {Beers}, T.~C., {Wilhelm}, R., \& {Lee}, Y.~S. 2006, ApJL, 638, L97

\bibitem[{{Tripp} {et~al.}(2003){Tripp}, {Wakker}, {Jenkins}, {Bowers},
  {Danks}, {Green}, {Heap}, {Joseph}, {Kaiser}, {Linsky}, \&
  {Woodgate}}]{tripp-etal-03-ComplexC}
{Tripp}, T.~M., {Wakker}, B.~P., {Jenkins}, E.~B., {Bowers}, C.~W., {Danks},
  A.~C., {Green}, R.~F., {Heap}, S.~R., {Joseph}, C.~L., {Kaiser}, M.~E.,
  {Linsky}, J.~L., \& {Woodgate}, B.~E. 2003, \aj, 125, 3122

\bibitem[{{Tufte} {et~al.}(1998){Tufte}, {Reynolds}, \&
  {Haffner}}]{tufte-etal-98-WHAM}
{Tufte}, S.~L., {Reynolds}, R.~J., \& {Haffner}, L.~M. 1998, \apj, 504, 773

\bibitem[{{van Woerden} {et~al.}(1999){van Woerden}, {Schwarz}, {Peletier},
  {Wakker}, \& {Kalberla}}]{vanwoerden-etal-99-nature}
{van Woerden}, H., {Schwarz}, U.~J., {Peletier}, R.~F., {Wakker}, B.~P., \&
  {Kalberla}, P.~M.~W. 1999, \nat, 400, 138

\bibitem[{{Vogt} {et~al.}(1994){Vogt}, {Allen}, {Bigelow}, {Bresee}, {Brown},
  {Cantrall}, {Conrad}, {Couture}, {Delaney}, {Epps}, {Hilyard}, {Hilyard},
  {Horn}, {Jern}, {Kanto}, {Keane}, {Kibrick}, {Lewis}, {Osborne},
  {Pardeilhan}, {Pfister}, {Ricketts}, {Robinson}, {Stover}, {Tucker}, {Ward},
  \& {Wei}}]{vogt-etal-94-HIRES}
{Vogt}, S.~S., {Allen}, S.~L., {Bigelow}, B.~C., {Bresee}, L., {Brown}, B.,
  {Cantrall}, T., {Conrad}, A., {Couture}, M., {Delaney}, C., {Epps}, H.~W.,
  {Hilyard}, D., {Hilyard}, D.~F., {Horn}, E., {Jern}, N., {Kanto}, D.,
  {Keane}, M.~J., {Kibrick}, R.~I., {Lewis}, J.~W., {Osborne}, J.,
  {Pardeilhan}, G.~H., {Pfister}, T., {Ricketts}, T., {Robinson}, L.~B.,
  {Stover}, R.~J., {Tucker}, D., {Ward}, J., \& {Wei}, M.~Z. 1994, in Presented
  at the Society of Photo-Optical Instrumentation Engineers (SPIE) Conference,
  Vol. 2198, Proc. SPIE Instrumentation in Astronomy VIII, David L. Crawford;
  Eric R. Craine; Eds., Volume 2198, p. 362, ed. D.~L. {Crawford} \& E.~R.
  {Craine}, 362--+

\bibitem[{{Wakker}(2001)}]{wakker-01-distance-metallicity}
{Wakker}, B.~P. 2001, \apjs, 136, 463

\bibitem[{{Wakker} {et~al.}(1999){Wakker}, {Howk}, {Savage}, {van Woerden},
  {Tufte}, {Schwarz}, {Benjamin}, {Reynolds}, {Peletier}, \&
  {Kalberla}}]{wakker-etal-99-nature}
{Wakker}, B.~P., {Howk}, J.~C., {Savage}, B.~D., {van Woerden}, H., {Tufte},
  S.~L., {Schwarz}, U.~J., {Benjamin}, R., {Reynolds}, R.~J., {Peletier},
  R.~F., \& {Kalberla}, P.~M.~W. 1999, \nat, 402, 388

\bibitem[{{Wakker} {et~al.}(2001){Wakker}, {Kalberla}, {van Woerden}, {de
  Boer}, \& {Putman}}]{wakker-etal-01-HI-spectra}
{Wakker}, B.~P., {Kalberla}, P.~M.~W., {van Woerden}, H., {de Boer}, K.~S., \&
  {Putman}, M.~E. 2001, \apjs, 136, 537

\bibitem[{{Wakker} {et~al.}(2002){Wakker}, {Oosterloo}, \&
  {Putman}}]{wakker-oosterloo-putman-02-fine-structure}
{Wakker}, B.~P., {Oosterloo}, T.~A., \& {Putman}, M.~E. 2002, \aj, 123, 1953

\bibitem[{{Wakker} \& {van Woerden}(1991)}]{wakker-vanwoerden-91-catalogue}
{Wakker}, B.~P. \& {van Woerden}, H. 1991, \aap, 250, 509

\bibitem[{{Wakker} {et~al.}(1996){Wakker}, {van Woerden}, {Schwartz},
  {Peletier}, \& {Douglas}}]{wakker-etal-96-ComplexC-CaII-abundance}
{Wakker}, B.~P., {van Woerden}, H., {Schwartz}, U.~J., {Peletier}, R.~F., \&
  {Douglas}, N.~G. 1996, \aap, 306, L25+

\bibitem[{{Wakker} {et~al.}(2007{\natexlab{a}}){Wakker}, {York}, {Howk},
  {Barentine}, {Wilhelm}, {Peletier}, {van Woerden}, {Beers}, {Ivezi{\'c}},
  {Richter}, \& {Schwarz}}]{wakker-etal-07-ComplexC}
{Wakker}, B.~P., {York}, D.~G., {Howk}, J.~C., {Barentine}, J.~C., {Wilhelm},
  R., {Peletier}, R.~F., {van Woerden}, H., {Beers}, T.~C., {Ivezi{\'c}}, {\v
  Z}., {Richter}, P., \& {Schwarz}, U.~J. 2007{\natexlab{a}}, \apjl, 670, L113

\bibitem[{{Wakker} {et~al.}(2007{\natexlab{b}}){Wakker}, {York}, {Wilhelm},
  {Barentine}, {Richter}, {Beers}, {Ivezi{\'c}}, \&
  {Howk}}]{wakker-etal-07-I-distances}
{Wakker}, B.~P., {York}, D.~G., {Wilhelm}, R., {Barentine}, J.~C., {Richter},
  P., {Beers}, T.~C., {Ivezi{\'c}}, {\v Z}., \& {Howk}, J. 2007{\natexlab{b}},
  \apj\ submitted (arXiv:0709.1926)

\bibitem[{{Westmeier} {et~al.}(2005){Westmeier}, {Braun}, \&
  {Thilker}}]{westmeier-etal-05-M31}
{Westmeier}, T., {Braun}, R., \& {Thilker}, D. 2005, \aap, 436, 101

\bibitem[{{Wilhelm} {et~al.}(1999){Wilhelm}, {Beers}, \&
  {Gray}}]{wilhelm-etal-99a-classification}
{Wilhelm}, R., {Beers}, T.~C., \& {Gray}, R.~O. 1999, \aj, 117, 2308

\bibitem[{{Willman} {et~al.}(2005{\natexlab{a}}){Willman}, {Blanton}, {West},
  {Dalcanton}, {Hogg}, {Schneider}, {Wherry}, {Yanny}, \&
  {Brinkmann}}]{willman-etal-05-willmanI}
{Willman}, B., {Blanton}, M.~R., {West}, A.~A., {Dalcanton}, J.~J., {Hogg},
  D.~W., {Schneider}, D.~P., {Wherry}, N., {Yanny}, B., \& {Brinkmann}, J.
  2005{\natexlab{a}}, \aj, 129, 2692

\bibitem[{{Willman} {et~al.}(2002){Willman}, {Dalcanton}, {Ivezi{\'c}},
  {Schneider}, \& {York}}]{willman-etal-02-stellar-content}
{Willman}, B., {Dalcanton}, J., {Ivezi{\'c}}, {\v Z}., {Schneider}, D.~P., \&
  {York}, D.~G. 2002, \aj, 124, 2600

\bibitem[{{Willman} {et~al.}(2005{\natexlab{b}}){Willman}, {Dalcanton},
  {Martinez-Delgado}, {West}, {Blanton}, {Hogg}, {Barentine}, {Brewington},
  {Harvanek}, {Kleinman}, {Krzesinski}, {Long}, {Neilsen}, {Nitta}, \&
  {Snedden}}]{willman-etal-05-UMa}
{Willman}, B., {Dalcanton}, J.~J., {Martinez-Delgado}, D., {West}, A.~A.,
  {Blanton}, M.~R., {Hogg}, D.~W., {Barentine}, J.~C., {Brewington}, H.~J.,
  {Harvanek}, M., {Kleinman}, S.~J., {Krzesinski}, J., {Long}, D., {Neilsen},
  Jr., E.~H., {Nitta}, A., \& {Snedden}, S.~A. 2005{\natexlab{b}}, \apjl, 626,
  L85

\end{thebibliography}
\bibliographystyle{apj}

\clearpage


\end{document}